# Cadherin & Claudin: It takes two to tango

Julie Gavard[1] and J. Silvio Gutkind[2]

[1] Institut Cochin, CNRS, UMR8104, INSERM, U567, Université Paris Descartes, 22 rue Méchain, 75014, Paris, FRANCE
[2] Oral and Pharyngeal Cancer Branch, National Institute of Dental and Craniofacial Branch, National Institutes of Health, Bldg 30, Room 211, Bethesda MD20892, USA

**The endothelial barrier function requires the adhesive activity of VE-cadherin and claudin-5, two key components of adherens and tight endothelial junctions, respectively. Emerging evidence that VE-cadherin controls claudin-5 expression by preventing the nuclear accumulation of FoxO1 and β-catenin, which repress the claudin-5 promoter, provides a crosstalk between these endothelial junctional structures.**

The formation of adhesive structures between adjacent cells, including adherens and tight junctions, contribute to the establishment of cell polarity, differentiation and survival, and is ultimately required for the maintenance of the tissue integrity. Cadherins, the main constituent of the adherens junctions, belong to a conserved family of adhesion molecules that provide a molecular bond between cells and link the plasma membrane with the intracellular actin cytoskeleton through catenin family proteins[1]. Tight junctions form a dense ultra-structure organization, observable by electron microscopy, which involves numerous adhesive molecules, including occludin, junctional adhesion molecules (JAMs), and the claudin family of tetraspan transmembrane proteins, as well as and intracellular adapters, such as ZO-1 and ZO-2[2]. Whereas in epithelial cells, tight junctions are often located apically with respect to adherens junctions, in other cells, such as in endothelial cells, the adherens and tight junctions are intermingled throughout cell-cell contact areas[3]. The formation, maintenance, and remodelling of the intercellular contacts requires a functional interaction between these two adhesive structures. For example, the barrier function of the endothelium requires the adhesive activity of VE-cadherin and claudin-5, which are key components of the adherens and tight endothelial junctions, respectively. Surprisingly, this functional relationship involves also the direct control of the expression of claudin-5 by VE-cadherin. Indeed, as shown on page xxx of this issue by Taddei *et al*, VE-cadherin-mediated adhesion enables claudin-5 expression in endothelial cells by preventing the nuclear accumulation of two transcriptional regulators, FoxO1 and β-catenin, which repress claudin-5 expression when VE-cadherin is absent or non-functional (**Fig. 1**). These findings clearly place VE-cadherin upstream of claudin-5 in the establishment, maturation and maintenance of endothelial cell-cell junctions and further suggest that any changes in VE-cadherin adhesive properties will impact at multiple levels on the endothelial barrier function.

At the molecular level, Taddei *et al.* observed that VE-cadherin adhesion triggers a sustained activation of the PI-3 kinase/AKT pathway, and the subsequent phosphorylation of FoxO1, a forkhead repressor transcription factor, which results in its cytosolic localization, thus preventing FoxO1 binding to its DNA targets within the nuclear compartment. In the absence of VE-cadherin at the cell surface, non-phosphorylated FoxO1 constitutively accumulates within the nucleus thereby repressing the claudin-5 promoter by acting directly on two FoxO1 response elements (**Fig. 1**). Strikingly, β-catenin is required for FoxO1 repression of claudin-5 gene expression, most likely through the stabilization of FoxO1 interactions to its DNA recognition sites. This depends on the T-cell factor (TCF), which is constitutively associated to the claudin-5 promoter. Further work is still required to understand fully how FoxO1 phosphorylation modulates its localization, interactions with β-catenin and TCF, and its transcriptional repressive activity.

The role of cadherins in the control of gene expression has been often associated to their negative effect on the canonical Wnt-initiated pathway, by a mechanism that involves the retention of β-catenin at the plasma membrane upon binding to clustered cadherins, which therefore limits the pool of free β-catenin available for nuclear shuttling[4]. The mechanism identified here in endothelial cells is quite reminiscent to that recently described in neuronal cells, in which Slit stimulation of its receptor, Robo, triggers the rapid loss of N-cadherin dependent adhesion, which leads to the release of β-catenin from adhesion complexes and its nuclear accumulation and activation of β-catenin/TCF target genes[5]. In light of these studies, it will be intriguing to determine whether FoxO1 might also participate in the Slit/Robo long-term remodelling of cell adhesion in axon guidance, and how the partition of β-catenin between FoxO and TCF is integrated in response to canonical Wnt signalling. In this regard, β-catenin's interaction with FoxO has been proposed to compete with TCF/β-catenin-dependent activation of canonical Wnt-target genes[6], while *Taddei et al.* suggest that TCF is required for β-catenin/FoxO1 stabilization on the claudin-5 promoter. This raises the question as to whether Wnt-regulated target genes are also affected by VE-cadherin adhesion, either directly by β-catenin trapping, and/or by FoxO competition. While VE-cadherin can be considered as a multifunctional adhesion receptor modulating endothelial biology, the emerging picture is that VE-cadherin may also dictate endothelial cell fate through the regulation of intracellular signalling circuitries controlling gene transcription *via* multiple DNA-binding proteins including, at least, β-catenin, TCF and FoxO1.

The nature of the adhesion molecules, such as VE-cadherin and claudin-5, distinguish endothelial junctions, as their pattern of expression is highly restricted to the endothelial lineage and their organization is quite distinct from more common junctions, such as those found in epithelial cells[3]. Their functions are non-redundant since, for instance, claudin-5 is the major claudin identified in normal endothelial cells while multiple claudins can be found at the surface of epithelial cells. Interestingly, while claudin-5 knockout mice develop normally but have a defective blood brain barrier function and die shortly after birth[7], VE-cadherin knockout mice are embryonic lethal and exhibit multiple severe defects during developmental angiogenesis [8], suggesting that VE-cadherin function may go far beyond promoting cell-cell adhesion. For example, VE-cadherin can modulate VEGF receptor signalling, which transduces the pro-angiogenic and pro-permeability effects of VEGF[8]. Indeed, many of the biological responses attributed to VE-cadherin appear intimately linked to VEGF signalling[9]. The work by Taddei et al. identifies a unique mechanism by which VE-cadherin can control claudin-5 expression, independently of VEGF/VEGF receptor downstream signalling, as blocking either VEGF or the activity of its receptor does not alter VE-cadherin control over claudin-5 gene expression. The molecular mechanism reported here supports a permissive role for VE-cadherin clustering on claudin-5 expression, and one can envision that VE-cadherin might act as key regulator of endothelial cell lineage, as ultimately claudin-5 expression is restricted to VE-cadherin expressing cells. However, more studies are required now to fully understand whether VE-cadherin can orchestrate the activation of an endothelial maturation program and/or the repression of a non-endothelial phenotype.

In addition to angiogenesis, there is a plethora of physiological and pathological conditions where the endothelial barrier integrity is compromised. Hierarchically, VE-cadherin is major in controlling the endothelial barrier function. For example, VE-cadherin endocytosis and uncoupling from catenin-associated proteins have been proposed to collectively contribute to the disruption of endothelial cell-cell junctions in response to VEGF[10, 11]. In light of the study published in page xxx of this issue, inhibition of VE-cadherin expression level, its adhesive ability, or its plasma membrane availability might provoke β-catenin and FoxO1 nuclear accumulation, and therefore the repression of claudin-5 transcription (**Fig. 1**). Thus, while the acute VEGF stimulation could induce an increase in vascular permeability by a reversible disruption of VE-cadherin adhesion and the subsequent disorganization of the tight junctions[10], the chronic exposure to VEGF or other pro-permeability factors might affect severely the endothelial barrier, as the turnover and the expression of claudin-5 could be compromised in the absence of functional VE-cadherin adhesion. The regulation of claudin-5 expression by VE-cadherin points to an intimate crosstalk between tight and adherens junctions. However, this crosstalk is bidirectional since tight junctional molecules, such as JAMs may regulate cadherin-mediated adhesion in endothelial cells[12].

Beside VEGF, reactive oxygen species can increase vascular permeability through a Rac-dependent mechanism, associated with a loss of endothelial cell-cell junctions[13]. Interestingly, recent work in C. elegans suggests that oxygen stress signalling depends on a functional interaction between FoxO and β-catenin[14]. The novel findings reported on page xxx allow us to speculate that reactive oxygen species might impact endothelial biology by an early action on VE-cadherin adhesion through a Rac signalling nexus and later by downregulating claudin-5 expression (**Fig. 1**). Overall, this two-pronged control of endothelial junctions by VE-cadherin function may contribute to the pathological disruption of the vascular wall, which is observed in variety of diseases such as in tumour-induced angiogenesis, stroke, myocardial infarction, inflammation, allergy, and macular degeneration, to name a few. It is also noteworthy that the severe blood brain barrier phenotype of claudin-5 knockout mice has elicited the attention on claudin-5 in search for treatments for diseases associated with aberrant blood brain barrier function, or to enhance the efficiency of drug delivery to the brain[7]. The signalling pathway characterized here may now offer the opportunity to reconsider VE-cadherin as a molecular target for the treatment of central nervous system diseases. For instance, brain stroke morbidity is associated with neural cell damage caused by the breakdown of the blood brain barrier, and the subsequent exudation of plasma components and oedema. In this regard, blocking VEGF pro-permeability signalling had successfully permitted to limit the size of the lesion in mouse models for brain ischemia[15]. We can now anticipate that restoring VE-cadherin adhesion and function might not only reduce vascular leakage, perhaps more effectively than VEGF antagonists, as it may also help rescuing the blood brain barrier function by providing a positive feedback on claudin-5 expression.

Vascular leakage and loss of the endothelial barrier integrity is a hallmark of pathological angiogenesis. Substantial progresses had been recently made in understanding the molecular mechanisms regulating endothelial cell-cell junctions. The direct impact of VE-cadherin on the regulation of expression of the tight junction protein claudin-5 provides now a further rationale for the development of therapeutic approaches targeting VE-cadherin and its downstream molecules for vascular "normalization" in many human diseases that involve aberrant angiogenesis and vascular leakage.

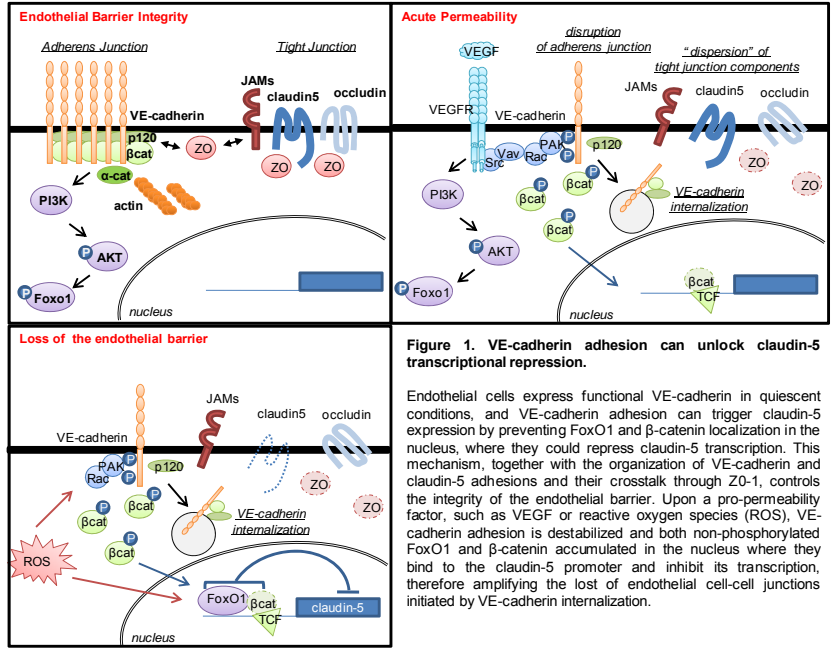

**Figure 1. VE-cadherin adhesion can unlock claudin-5 transcriptional repression.**

Endothelial cells express functional VE-cadherin in quiescent conditions, and VE-cadherin adhesion can trigger claudin-5 expression by preventing FoxO1 and β-catenin localization in the nucleus, where they could repress claudin-5 transcription. This mechanism, together with the organization of VE-cadherin and claudin-5 adhesions and their crosstalk through Z0-1, controls the integrity of the endothelial barrier. Upon a pro-permeability factor, such as VEGF or reactive oxygen species (ROS), VE-cadherin adhesion is destabilized and both non-phosphorylated FoxO1 and β-catenin accumulated in the nucleus where they bind to the claudin-5 promoter and inhibit its transcription, therefore amplifying the lost of endothelial cell-cell junctions initiated by VE-cadherin internalization.